\documentclass[aps,preprint,superscriptaddress]{revtex4}

\usepackage[english]{babel}
\usepackage{amsmath}
\usepackage{graphicx}
\usepackage{hyperref}
\usepackage[usenames,dvipsnames]{xcolor}
\hypersetup{colorlinks=true, citecolor=Plum, linkcolor=Plum,
urlcolor=Plum}

\begin{document}

\title{Ten years of gravitational-wave astronomy}
\author{Emanuele Berti}
\affiliation{William H. Miller III Department of Physics and Astronomy, Johns Hopkins University, Baltimore, Maryland 21218, USA}

\begin{abstract}
  Ten years ago humankind achieved the first direct observation of gravitational waves. I give some personal recollections of that first detection. I also present an incomplete summary of what we have learned since then, and some speculations on what we may learn in the future. 
\end{abstract}

\maketitle

There are dates that remain in our memories, for good or bad.
Every adult who saw the Twin Towers collapse on September 11, 2001 remembers where they were on that day.
I was in an overcrowded Ph.D. student room at the Department of Physics in Rome, working on some problem in black hole perturbation theory. I had visited the US for the first time in November of the previous year. Like many tourists, I had decided to visit the top floor of the Twin Towers (the ``Top of the World'') to admire the view. That day I opened the news and saw the planes crashing into the towers, then the internet connection crashed as well -- too much traffic. Small crowds gathered outside the physics department building to hear the devastating news from the radio.

September 14, 2015 was a similarly memorable date, but for a happy reason. While the LIGO detectors were operating in ``engineering mode'' -- the formal  ``observing'' phase was only scheduled to start three days later -- humanity caught the first gravitational wave signal. The waves emitted by two black holes that merged approximately 1.3 billion years ago, when life on Earth was primarily microbial and multicellular life as we know it was just beginning to emerge, produced a displacement in the LIGO mirrors smaller than the size of an atomic nucleus, but large enough to be detected~\cite{LIGOScientific:2016aoc}.

Most scientists who were not members of the LIGO-Virgo Collaboration (me included) learned about the detection weeks or months later. As in the case of the Twin Towers, I would bet that most of them could tell you where they were when they first heard the news.  At the time I had been working for some years with Chris Belczynski's student Michal Dominik, who came to visit me in Mississippi. We were using their population synthesis models to predict binary black hole merger rates. We had come to the conclusion that the upcoming LIGO detector observing run would have to observe {\em some} binary mergers, unless the models were wrong~\cite{Dominik:2014yma} (see~\cite{Mandel:2021smh} for a review). Then there were rumors on social media. Our LIGO-Virgo colleagues started behaving in secretive ways. Chris and some of our common LIGO friends became unusually excited. Something was definitely going on.

My son was two years old at the time and my daughter was born in December 2015,
so I had many sleepless nights in those days. On January 20, 2016, I did not sleep for a different reason. That day I got an e-mail from Jessica Thomas with subject {\em Urgent and confidential request from the editors of Physics and PRL}.
At first I thought they would ask me to referee {\em that} paper. Instead the email said:
\noindent
{\em ``I'm the Editor of \href{https://physics.aps.org}{\underline{Physics}}, a web publication that features news and commentary about papers in PRL and the Physical Review. I'm writing to you, in confidence, because my colleague at PRL, Robert Garisto, has alerted me to a very exciting paper related to the search for gravitational waves that the journal expects to receive tomorrow. Assuming this paper passes the scrutiny of peer review, we will definitely want to highlight the work in Physics with a Viewpoint commentary. Would you be interested in writing this commentary for us?''} Hell yes, I was!

The next day I checked my email frantically, but the ``very exciting paper'' was not there.
I finally found it in my inbox just a few minutes before the daycare pickup time. My jaw dropped when I saw the now famous plot of the measured signal in the Hanford and Livingston detectors (reproduced in Figure 2 of the \href{https://physics.aps.org/articles/v9/17}{\underline{Viewpoint}}~\cite{Berti:2016qkk}), and I almost missed my daycare pickup. The ``very exciting paper'' was published on February 11, 2016. That day the PRL website crashed -- too much traffic. The Nobel committee took notice. The way in which we observe the Universe changed forever.

In the intervening decade we have learned a lot about the astrophysics of compact binaries and about gravity itself -- both from individual events, and from the population of merging binaries as a whole.

\section*{An incomplete list of notable individual events}

In the past ten years, the LIGO-Virgo-KAGRA network has observed many notable individual events.
Each of them advanced -- or challenged -- our understanding of the astrophysics of compact objects.
This is an incomplete list of some of those events and what they taught us.

\noindent
{\bf GW150914}~\cite{LIGOScientific:2016aoc} was the first direct detection of gravitational waves. It also proved that astrophysical black holes of mass $\sim 30\,M_\odot$, much heavier than the electromagnetically inferred stellar mass black holes observed up to that point, exist and merge.

\noindent
{\bf GW151226}~\cite{LIGOScientific:2016sjg}, the ``Boxing Day event,'' is a relatively long signal ($\sim 55$ cycles). It gave us the first gravitational-wave measurement of black hole spin (at least one of the components has spin $>0.2$) and of nonzero spin misalignment. Kinematical arguments imply that the merger remnant received a natal kick $v_{\rm kick}\geq 50\,{\rm km}/{\rm s}$~\cite{OShaughnessy:2017eks}.

\noindent
{\bf GW170817}~\cite{LIGOScientific:2017vwq,LIGOScientific:2017ync} was the first observation of a binary neutron star merger. Its association with the $\gamma$-ray burst GRB 170817A, detected by Fermi-GBM 1.7\,s after the coalescence, provided the first direct evidence of a link between binary neutron star mergers and short $\gamma$-ray bursts. Transient counterparts identified in the UV, optical, near-infrared, X-ray and radio confirmed that the event was produced by the merger of two neutron stars, followed by a short $\gamma$-ray burst and a kilonova/macronova powered by the radioactive decay of r-process nuclei synthesized in the ejecta. These multi-messenger observations also proved that heavy elements, such as lead and gold, are created in these collisions.

\noindent {\bf GW190412}~\cite{LIGOScientific:2020stg} proved that black hole mergers can have very asymmetric masses (here $\sim 30M_\odot$ and $\sim 8 M_\odot$). This system is hard to explain by isolated evolution, but also via dynamical formation, and it lent further support to the idea that some black hole mergers can occur in nuclear star clusters or AGNs.

\noindent {\bf GW190425}~\cite{LIGOScientific:2020aai} was consistent with the binary components being neutron stars, although the possibility that one or both components of the system are black holes cannot be ruled out from gravitational-wave data. If the binary components are indeed neutron stars, this event suggests that neutron star binaries could be systematically more massive than the known Galactic neutron star population.

\noindent
{\bf GW190521}~\cite{LIGOScientific:2020iuh} was a merger of unusually heavy black holes: the primary black hole mass is within the so called ``upper mass gap'' produced by (pulsational) pair-instability supernova (PISN/PPISN) processes, where black holes should not form from stellar collapse, and the remnant mass is $\sim 142\,M_\odot$. This event proved that the upper mass gap can be populated, and that intermediate-mass black holes exist in the Universe.

\noindent
{\bf GW190814}~\cite{LIGOScientific:2020zkf}, with its $(23+2.6)\,M_\odot$ components, was the event with the most unequal mass ratio yet measured. It contains either the lightest black hole or the heaviest neutron star ever discovered in a double compact-object system. The dimensionless spin of the primary black hole is tightly constrained to $\leq 0.07$. Astrophysical models predict that binaries with mass ratios similar to GW190814 can form through several channels, but are unlikely to have formed in globular clusters. The combination of mass ratio, component masses, and the inferred merger rate for this event challenges all current models for the formation and mass distribution of compact binaries.

\noindent
{\bf GW200105\_162426} and {\bf GW200115\_042309}~\cite{LIGOScientific:2021qlt} provided evidence that neutron stars can merge with black holes, and the first constraints on the rate of these mergers.

\noindent
{\bf GW200129} was claimed to provide evidence for orbital precession~\cite{Hannam:2021pit} (but see~\cite{Payne:2022spz} for caveats related to data quality) and the first identification of a large kick velocity for the remnant, $v_{\rm kick}\gtrsim 698$\,km/s at 90\% credibility~\cite{Varma:2022pld}.

\noindent
{\bf GW230529}~\cite{LIGOScientific:2024elc}, with component masses in the range  $[2.5-4.5]\,M_\odot + [1.2–2.0]\,M_\odot$ at 90\% confidence level, showed that compact objects can populate the hypothetical ``lower mass gap'' that could separate the heaviest neutron stars from the lightest black holes -- i.e., the primary has mass $<5\,M_\odot$ at 90\% credibility. Given present estimates of the maximum neutron star mass, the most probable interpretation of the source is a neutron star-black hole coalescence, where the black hole has mass between the most massive neutron stars and the least massive black holes observed in the Galaxy.

\noindent 
{\bf GW231123}~\cite{LIGOScientific:2025rsn} has the largest observed component masses $\sim (140+100)\,M_\odot$ and very large component spins, with an inferred primary spin $\chi_1\sim 0.9$. Some properties of GW231123 are subject to large systematic uncertainties, as the inferred parameters vary when using different waveform models. The primary black hole lies within {\em or above} the upper mass gap at $[60-130]\,M_\odot$ where black holes should be rare due to the PISN/PPISN mechanism, while the secondary spans the gap. This event suggests that at least some black holes should form from channels beyond standard stellar collapse, and that intermediate mass black holes of mass $\sim 200\,M_\odot$ could form through hierarchical mergers~\cite{Gerosa:2017kvu,Fishbach:2017dwv,Gerosa:2021mno}.

\noindent 
{\bf GW250114}~\cite{LIGOScientific:2025epi} is a ``twin'' of GW150914, but thanks to the impressive experimental progress in the intervening decade, it is the loudest signal observed so far, with a network signal-to-noise ratio of 80 in the two LIGO detectors. The post-merger data are consistent with the dominant quadrupolar mode of the radiation and its first overtone, which allows for the first test of the Kerr black hole nature of the merger remnant based on ``black hole spectroscopy''~\cite{Detweiler:1980gk,Dreyer:2003bv,Berti:2005ys,Berti:2009kk,Berti:2025hly}. By measuring the mass and spin of the progenitors and of the remnant with a range of analyses that exclude up to five of the strongest merger cycles, this event also confirms the Bekenstein-Hawking area law (also known as the ``second law of black hole mechanics''), which states that the total area of the black hole's event horizons is proportional to their entropy and cannot decrease with time.

\section*{Astrophysical population properties}

At the end of
the first part of the fourth observing run (O4a)
we have a catalog of $\sim 150$ merger detections~\cite{LIGOScientific:2025pvj}. 
The neutron star-neutron star, neutron star-black hole, and black hole-black hole merger rates are estimated to be in the range
$[7.6,\,250]\,{\rm Gpc}^{-3}\,{\rm yr}^{-1}$,
$[9.1,\,84]\,{\rm Gpc}^{-3}\,{\rm yr}^{-1}$, and
$[14,\,26]\,{\rm Gpc}^{-3}\,{\rm yr}^{-1}$,
respectively~\cite{LIGOScientific:2025pvj}.

The shape of the mass spectrum shows several interesting features, including possibly a peak around $10\,M_\odot$ and one around $35\,M_\odot$~\cite{LIGOScientific:2025pvj}; 
the nature of these peaks is debated, and it's unclear whether the second peak is related to PISN/PPISN instabilities~\cite{Farah:2023vsc,Roy:2025ktr}.

These observations prompt lots of questions that we may be able to answer as we measure more binaries with more sensitive interferometers and better waveform models (see e.g.~\cite{Callister:2024cdx} for a review):

\begin{itemize}

\item[(i)] What is our best interpretation of the observed binary black hole mass spectrum~\cite{Mandel:2025ngw}? How well do we actually observe the spectrum? What masses should merging black holes have, and can we reconcile gravitational-wave and electromagnetic measurements?
  
\item[(ii)] Are there correlations between the observed binary parameters? In particular, do more unequal-mass mergers have larger effective spins, as suggested by a statistical analysis of the data~\cite{Callister:2021fpo}? If so, why?

\item[(iii)] The observed black hole spins seem to be generally low. Is this a real physical effect, ot are they just poorly measured~\cite{Callister:2022qwb}?

\item[(iv)] How do the merger rates and the binary properties evolve with redshift~\cite{vanSon:2021zpk,Rinaldi:2023bbd,Lalleman:2025xcs}? Does the effective spin distribution broaden with redshift, as some observations seem to suggest -- and if so, why~\cite{Bavera:2022mef,Biscoveanu:2022qac,Ye:2024ypm}? Are there hints of spin-orbit resonances in the population~\cite{Varma:2021xbh}, and can they be used to infer the binary formation history~\cite{Gerosa:2013laa,Gerosa:2014kta,Kesden:2014sla,Gerosa:2015tea,Gerosa:2018wbw,Gangardt:2021lic,Gangardt:2022ltd}?

\end{itemize}

The ongoing fourth observing run will further increase the current sample of events. As we uncover more information about the compact binary population, some of these questions may find better answers.

\section*{Tests of gravity and of the nature of compact objects}

We are only beginning to use gravitational-wave observations to answer some of the big physics questions. Is general relativity the correct theory of gravity~\cite{Will:2014kxa,Berti:2015itd,Yunes:2025xwp}? Are we really sure that the compact objects observed by the LIGO-Virgo-KAGRA collaboration are consistent with the vacuum black hole solutions predicted by general relativity~\cite{Cardoso:2019rvt}?

There have been many speculations on possible observational signatures of quantum gravity and exotic compact objects, but all observations performed by the LIGO-Virgo-KAGRA collaboration so far are compatible with general relativity~\cite{LIGOScientific:2016lio,Yunes:2016jcc,LIGOScientific:2019fpa,LIGOScientific:2020tif,LIGOScientific:2021sio}.  ``Black hole spectroscopy'' tests~\cite{Detweiler:1980gk,Dreyer:2003bv,Berti:2005ys,Berti:2009kk}, which require the measurement of more than one oscillation mode, are still in their infancy~\cite{LIGOScientific:2025epi}. Better tests will require accurate models of nonlinearities in general relativity, and they must take into account several data analysis subtleties~\cite{Berti:2025hly}.

More pedantically, we need to do our homework and better understand gravitational-wave emission {\em within} general relativity. Our limited knowledge of gravitational waveforms is already affecting the astrophysical interpretation of certain events, such as GW190521~\cite{LIGOScientific:2020iuh} and GW231123~\cite{LIGOScientific:2025rsn}. Some events, like GW200129\_065458~\cite{Maggio:2022hre}, exhibit false violations of general relativity that have been attributed either to waveform systematics (such as mismodeling of spin precession) or to data-quality issues.

Better waveform models will be essential to test general relativity~\cite{Chandramouli:2024vhw} and to measure cosmological parameters~\cite{Dhani:2025xgt} in the near future. Potential causes that can lead to a false identification of a violation of general relativity in the era of next-generation detectors include detector noise, signal overlaps, gaps in the data, detector calibration, source model inaccuracy, missing physics in the source and in the underlying environment model, source misidentification, and mismodeling of the astrophysical population~\cite{Gupta:2024gun}.

\section*{Where are we going next?}

As in all branches of astronomy, our ability to test strong gravity, the nature of compact objects, and their astrophysical formation scenarios will depend on building a large network of sensitive detectors on the ground and in space. The science case of the Einstein Telescope~\cite{Abac:2025saz}, Cosmic Explorer~\cite{Evans:2023euw}, and space-based interferometers (such as LISA~\cite{LISA:2024hlh}, TianQin~\cite{TianQin:2015yph} or Taiji~\cite{Hu:2017mde}) is outstanding. A partial list of possibilities to expand the frequency and amplitude reach of current experiments includes high-frequency detectors~\cite{Ackley:2020atn,Aggarwal:2025noe}, future deciHz~\cite{Sedda:2019uro}, mHz~\cite{Baibhav:2019rsa} or $\mu$hz~\cite{Sesana:2019vho} detectors in space, lunar detectors~\cite{1990AIPC..202..188S,Cozzumbo:2023gzs,Jani:2020gnz,Ajith:2024mie}, and atom interferometry experiments such as MAGIS~\cite{MAGIS-100:2021etm}, AION~\cite{Badurina:2019hst} and AEDGE~\cite{AEDGE:2019nxb}.

Despite the difficult funding climate, the (gravitational) future is bright.

\section*{Acknowledgments}

E.B. is supported by NSF Grants No.~AST-2307146, PHY-2513337, PHY-090003, and PHY-20043, by NASA Grant No.~21-ATP21-0010, by John Templeton Foundation Grant No.~62840, by the Simons Foundation [MPS-SIP-00001698, E.B.], by the Simons Foundation International, and by Italian Ministry of Foreign Affairs and International Cooperation Grant No.~PGR01167.

\bibliographystyle{apsrev4-1}
\bibliography{references}

\end{document}